\documentclass[%
 reprint,
 amsmath,amssymb,
 aps,
aps,prb,reprint,twocolumn,superscriptaddress]{revtex4-2}

\usepackage{graphicx}
\usepackage{dcolumn}
\usepackage{bm}
\usepackage{amsmath}
\usepackage{float}
\usepackage{chngcntr}  
\usepackage[mathlines]{lineno}
\usepackage{gensymb}
\usepackage[utf8]{inputenc}

\begin{document}

\preprint{APS/123-QED}

\title{Agentic multi-fidelity learning of quasiparticle and excitonic properties}


\newcommand{\LANLT}{Theoretical Division, Los Alamos National Laboratory,
Los Alamos, New Mexico 87545, USA}
\newcommand{\CINT}{Center for Integrated Nanotechnologies, Materials Physics and
Applications Division, Los Alamos National Laboratory, Los Alamos,
New Mexico 87545, USA}

\author{Arnab Neogi}
\email{aneogi2@lanl.gov} %
\thanks{Corresponding author}
\affiliation{\LANLT}
\affiliation{\CINT}

\author{Aaron Forde}
\affiliation{\LANLT}

\author{Christopher A. Lane}
\affiliation{\LANLT}

\author{Sergei Tretiak}
\email{serg@lanl.gov}
\thanks{Corresponding author}
\affiliation{\LANLT}
\affiliation{\CINT}

\author{Jian\mbox{-}Xin Zhu}
\email{jxzhu@lanl.gov}
\thanks{Corresponding author}
\affiliation{\LANLT}
\affiliation{\CINT}

\date{\today}
\begin{abstract}

Many-body GW–Bethe–Salpeter equation calculations are essential for accurate simulations of electronic structure and optical properties in modern low-dimensional nanomaterials. However, these methods are computationally demanding and can exhibit localized numerical instabilities or convergence failures that are difficult to detect within high-throughput workflows. We introduce an agent-guided multi-fidelity framework for correcting GW--Bethe--Salpeter excited-state landscapes in strained MoS$_2$-WS$_2$ bilayers. Across stacking registries, strain branches and reciprocal-space samplings, the workflow identifies spike-like excursions, near-zero-gap collapse and cross-fidelity inconsistencies associated with fragile long-wavelength dielectric screening. A structural agent evaluates calculations by assigning confidence weights and selectively using a small number of high-accuracy reference points. Machine learning models then transfer information across related systems and apply Gaussian-process corrections to recover improved quasiparticle gaps and exciton binding energies, with calibrated uncertainty estimates. The approach corrects numerically induced artifacts without erasing physical strain dependence and substantially improves agreement with higher-fidelity references relative to a no-agent baseline. These results show that reliable surrogate learning for excited-state materials requires explicit diagnosis of numerical fragility, not direct interpolation of raw first-principles data points. The proposed framework is readily transferable to other optoelectronic nanomaterials characterized by strong quantum confinement, such as quantum dots, nanoribbons, layered two-dimensional semiconductors, and hybrid perovskite nanostructures.
\end{abstract}


\maketitle


\section{\label{sec:intro}INTRODUCTION}

Materials discovery is entering a regime in which predictive models, autonomous workflows, and large language model agents are expected to operate on increasingly large scientific datasets. This shift creates a basic requirement that the underlying data must be generated at sufficient scale and with sufficient reliability for learning. Ground-state density functional theory (DFT) has already enabled many high-throughput materials databases, where structural stability, total energies, and approximate electronic properties can be computed for large numbers of compounds \cite{kirklin2015open,horton2019high,jain2013commentary,horton2025accelerated,ghosh2020machine}. Excited-state materials data are much less developed \cite{veril2021questdb,westermayr2020machine}. Yet many of the quantities that determine optoelectronic performance, including quasiparticle bandgaps, optical excitation energies, exciton binding energies, oscillator strengths, and screening-dependent spectral trends, require many-body excited-state methods rather than ground-state calculations alone.

\begin{figure*}[!t]
\centering
\includegraphics[scale=0.5]{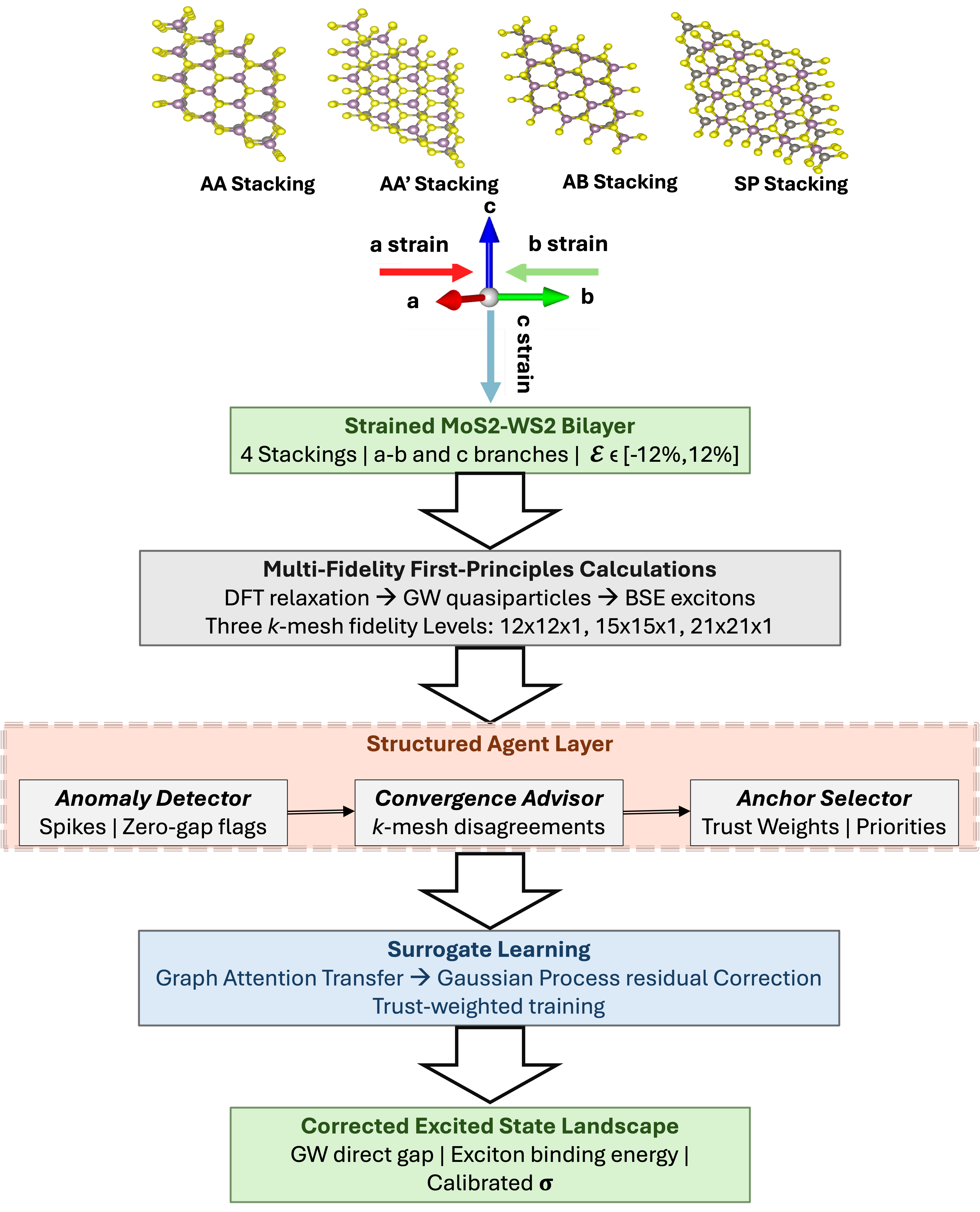}
\caption{\textbf{Agent-guided workflow for corrected strain-dependent GW-BSE properties.}
Schematic overview of the closed-loop workflow used to construct corrected excited-state data for strained MoS$_2$-WS$_2$ bilayers. The input design space consists of four stacking registries and two strain branches over $\epsilon \in [-12\%,12\%]$. For each configuration, multi-fidelity first-principles calculations are performed through DFT relaxation, GW quasiparticle calculations, and BSE exciton calculations using three $k$-mesh fidelities, $12\times12\times1$, $15\times15\times1$, and $21\times21\times1$. A structured agent layer then evaluates the resulting dataset through anomaly detection, convergence assessment, and anchor selection. Spike-like excursions, zero-gap flags, and $k$-mesh disagreements are converted into trust weights and anchor priorities, which determine how strongly each computed point contributes to subsequent learning. The trust-weighted data are used in a surrogate model combining graph-attention transfer with Gaussian-process residual correction. The final output is a corrected excited-state landscape with corrected GW direct gaps, exciton binding energies, and calibrated uncertainties.}
\label{Schematic_fig}
\end{figure*}

This bottleneck is particularly critical for low-dimensional semiconductors and van der Waals heterostructures. In transition-metal dichalcogenide bilayers, strain, stacking registry, and interlayer separation can strongly modify band alignment, interlayer hybridization, dielectric screening, exciton localization, and optical response \cite{manzeli20172d,mak2016photonics,peng2020strain, fan2022transition}. These sensitivities make strained heterobilayers promising platforms for excitonic engineering and strain-tunable optoelectronics \cite{ahn2017strain, leng2016gw}. They also make them difficult systems for predictive computation, because small structural changes can produce large changes in the excited-state landscape \cite{withers2015light,qi2023recent,chen2020chiral,du2021strain}. A useful dataset for this class of materials must therefore resolve not only a few representative structures, but many related configurations across stacking, strain, and reciprocal-space sampling.

The standard first-principles route to these quantities combines GW quasiparticle calculations with the Bethe-Salpeter equation (GW-BSE) \cite{aryasetiawan1998gw,hedin1965new, aulbur1999quasiparticle, cutkosky1954solutions, leng2016gw}. This approach is physically powerful, but it is not a simple extension of ordinary high-throughput DFT. A GW-BSE workflow is stage-wise \cite{ye2014probing,chaves2020bandgap} in the sense that the calculations must pass through ground-state relaxation, construction of a one-electron manifold, quasiparticle correction, dielectric screening, and solution of the electron-hole problem \cite{aryasetiawan1998gw, hedin1965new, aulbur1999quasiparticle}. Each stage introduces its own convergence requirements and possible failure modes. The simulation is computationally demanding, but cost is not the only challenge. In high-throughput excited-state calculations, the more difficult problem is often to determine whether a completed calculation has produced a trustworthy data point.

This distinction matters because many-body excited-state data can fail silently. A calculation may complete successfully and still produce a quasiparticle gap or excitonic quantity that is numerically fragile \cite{gan20222d,gopalan2022theoretical}. In quasi-two-dimensional systems, the long-wavelength dielectric response is especially delicate, and incomplete reciprocal-space convergence can propagate through the screened Coulomb interaction into quasiparticle corrections and excitonic observables \cite{zhang2023effect,gulans2014exciting,druppel2018electronic}. The resulting landscapes can contain  near-zero-gap collapse, cross-mesh inconsistencies, and artificial discontinuous strain trends. These artifacts are not equivalent to ordinary random noise. They are structured numerical failures that can be localized to specific strain windows, stacking registries, or reciprocal-space samplings \cite{varrassi2025automated,biswas2023py}. If such deficient data points are used directly for machine learning, a flexible surrogate can learn the numerical pathology along with the physical trend \cite{vinod2023multifidelity,yang2022multi,baum2026transfer}.

Recent work has begun to address different parts of this problem. Automated workflows for GW and GW-BSE calculations have improved the ability to launch and manage expensive excited-state computations \cite{varrassi2025automated,biswas2023py}. Multi-fidelity learning has shown that lower-cost calculations can help reduce the number of expensive high-fidelity excited-state calculations required for prediction \cite{vinod2023multifidelity,yang2022multi}. These advances do not fully solve the data-quality problem that appears in large, stage-wise GW-BSE datasets. Automation can produce more calculations, but it does not by itself decide which completed data points are reliable. Multi-fidelity learning can transfer information across levels of theory, but it can also transfer contaminated information if the training set mixes physically meaningful trends with numerically unstable data points. What is needed is a workflow that can build a usable training dataset from a large collection of heterogeneous excited-state calculations by identifying reliable evidence, reducing the influence of suspicious points, and selecting the most valuable higher-fidelity anchors.

Here we develop such a workflow for strained MoS$_2$-WS$_2$ bilayers \cite{pak2017strain,he2016strain}. We start with generating a multi-fidelity GW-BSE dataset across four stacking registries, two strain branches, and three reciprocal-space samplings, as illustrated in Fig.~\ref{Schematic_fig}. The dataset spans smooth regions, direct-to-indirect crossover regimes, incomplete workflow cases, and anomaly-prone strain windows. We then introduce a structured agent layer that evaluates physically motivated data-quality indicators, including spike scores, near-zero-gap flags, cross-mesh disagreement, stage-completion status, symmetry consistency, and numerical metadata. The agent assigns trust scores, recommended actions, training weights, and anchor priorities under a fixed schema. Its purpose is not to replace first-principles theory or to generate physical data points. Instead, it converts diagnostic information from the high-throughput workflow into decisions about how strongly each computed point should influence learning.
The agent-selected data are used to train a multi-fidelity surrogate for corrected excited-state landscapes. A graph-attention transfer model  \cite{louis2020graph, velivckovic2017graph, wang2019heterogeneous} learns strain-ordered relationships between lower and higher-fidelity calculations, while Gaussian-process residual correction \cite{heinze2018causal, hu2022causality, deringer2021gaussian, williams1995gaussian}  removes remaining structured errors and provides predictive uncertainty. The surrogate model development is described in Method Secs. H and I. The final outputs are corrected GW direct bandgap and exciton binding-energy curves as functions of strain. This design addresses two interconnected goals.
First, it distinguishes reliable training evidence from numerically fragile or incomplete outputs in a large set of stage-wise GW-BSE calculations that would be impractical to inspect manually. Second, it uses the resulting curated dataset to reconstruct expensive excited-state properties with a transfer-learning model that preserves physically meaningful strain dependence while correcting the localized numerical artifacts.
The broader message is that reliable data-driven excited-state materials design requires more than high-throughput execution of expensive first-principles calculations. It requires a closed loop in which simulation, diagnostic evaluation, agent-guided data selection, multi-fidelity learning, and uncertainty-aware correction operate together. In this framework, the central object is not the raw GW-BSE output alone, but a corrected and reliability-aware excited-state landscape suitable for surrogate prediction and future adaptive refinement.

\begin{figure*}[!t]
\centering
\includegraphics[scale=0.5]{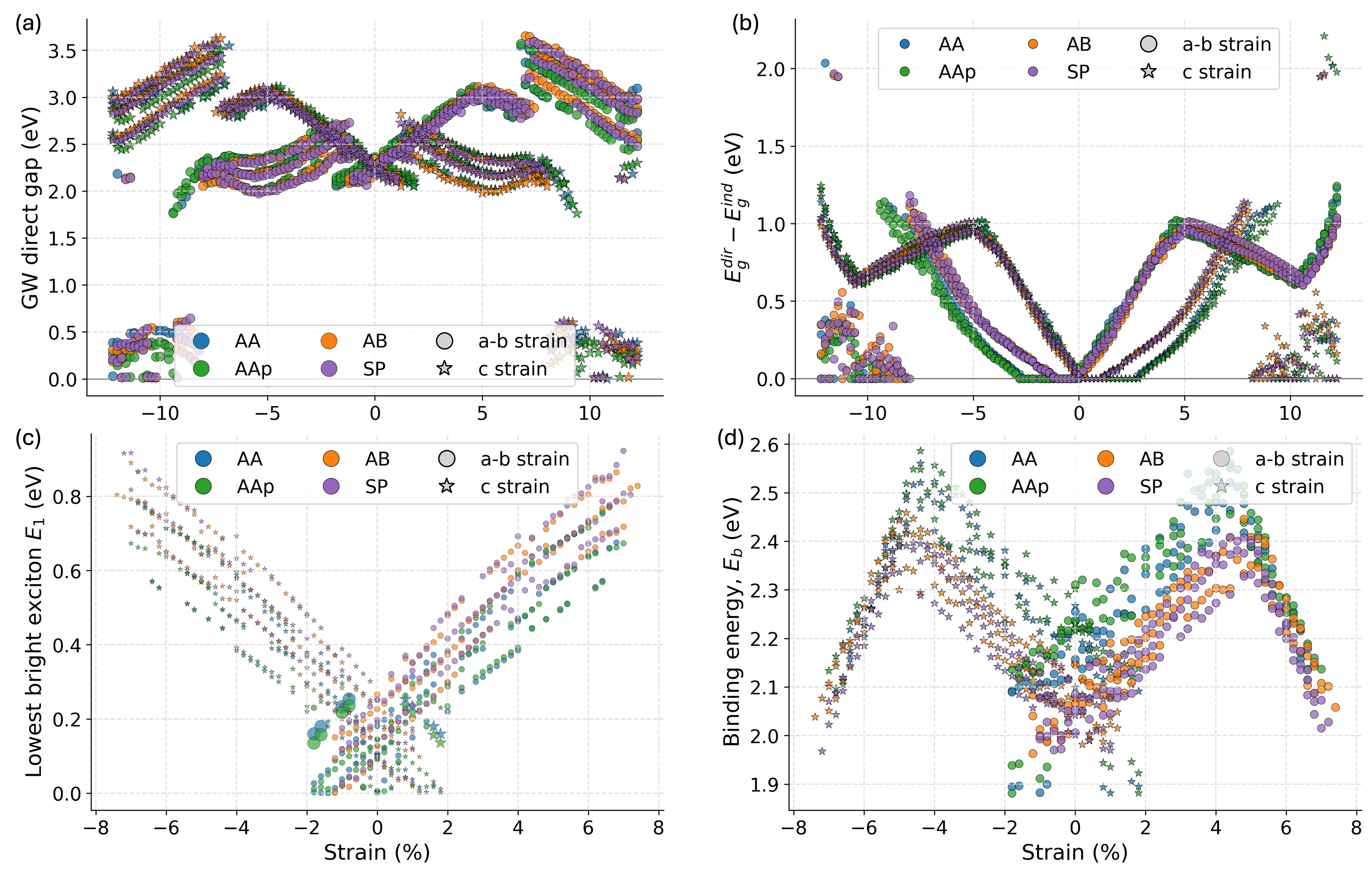}
\caption{\textbf{Raw multi-fidelity excited-state landscapes across strain.} Strain-dependent excited-state observables for the MoS$_2$-WS$_2$ bilayer dataset across stacking registries, strain branches, and reciprocal-space samplings. (a) GW direct bandgap. (b) Direct exciton binding energy, defined from the difference between the GW direct gap and the direct optical excitation energy. (c) Lowest-exciton binding energy, $E_b$. (d) Lowest bright exciton energy, $E_1$. Across broad regions of the strained design space, the raw landscapes evolve smoothly and reflect the expected sensitivity of band alignment, dielectric screening, and excitonic response to deformation. Selected stacking-strain environments, however, show abrupt excursions or spikes, near-zero-gap behavior, and discontinuous changes as shown in Fig. \ref{DFT_with_flags_fig}, that are inconsistent with the surrounding strain trends. These raw multi-fidelity landscapes therefore contain both physically meaningful strain dependence and localized numerical fragility, motivating anomaly-aware correction before surrogate learning.}\label{DFT_results_fig}
\end{figure*}

\section{Results}\label{sec:results}



\begin{figure*}[!t]
\centering
\includegraphics[scale=0.44]{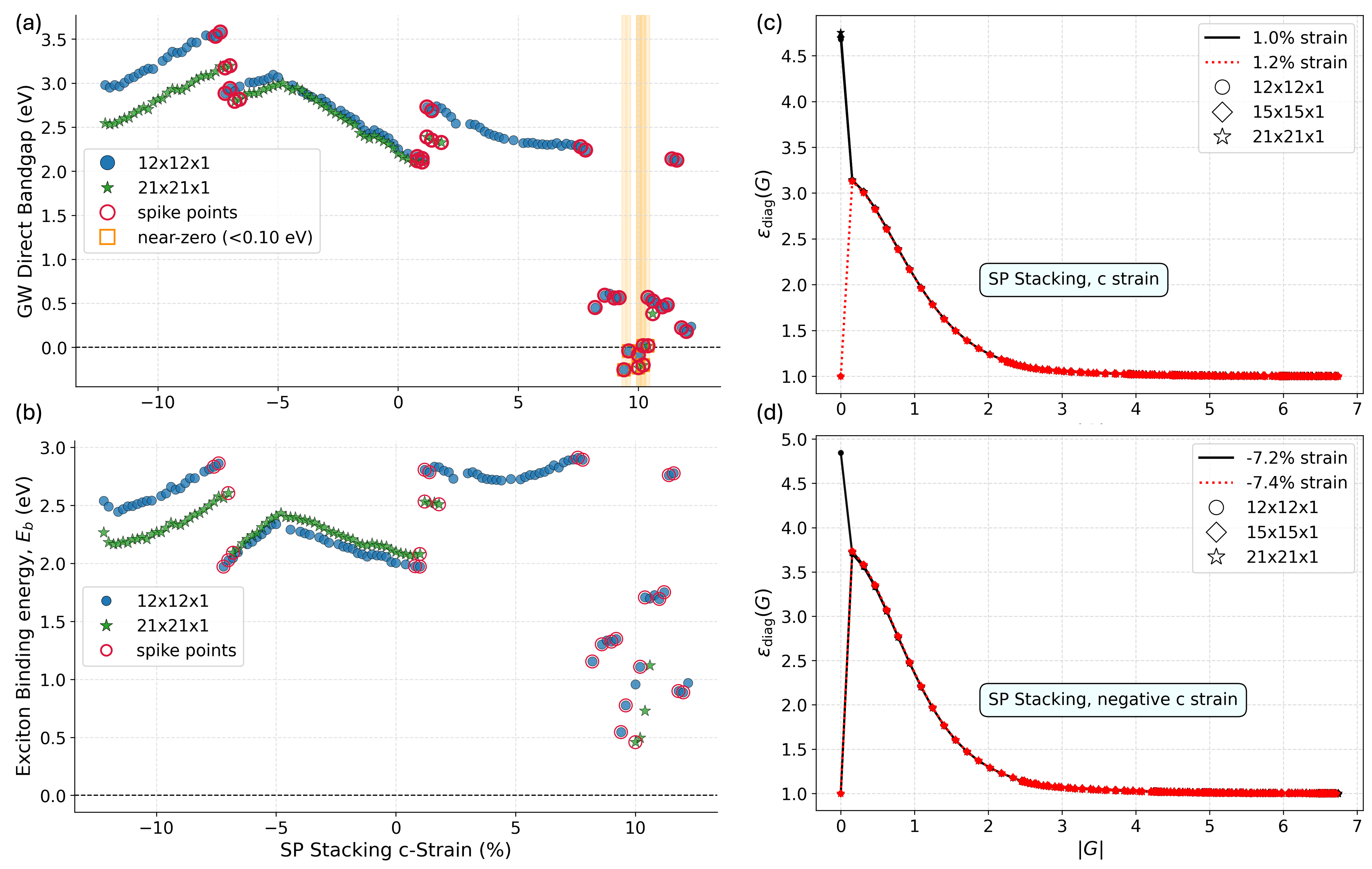}
\caption{\textbf{Representative anomaly-prone regime and its dielectric-screening origin.} (a) GW direct bandgap as a function of strain for SP stacking under the $c$-strain branch, comparing coarse and refined reciprocal-space samplings. Highlighted points identify localized spike anomalies and near-zero-gap configurations. (b) Corresponding exciton binding energy along the same structural branch, showing that the anomalous behavior appears simultaneously in quasiparticle and excitonic observables. (c,d) Diagonal microscopic dielectric screening, $\epsilon_{\mathrm{diag}}(G)$, for representative positive and negative $c$-strain configurations in the same anomaly-prone environment. The strong mesh sensitivity in the small-$|G|$ regime indicates incomplete convergence of the dielectric head and provides the physical origin of the spike-like anomalies observed in the coarse GW-BSE landscapes. Together, these panels show that the anomalous strain windows arise from localized numerical fragility in long-wavelength screening rather than from ordinary statistical noise, motivating trust-aware correction rather than naive interpolation through all computed points.}\label{DFT_with_flags_fig}
\end{figure*}

\begin{figure*}[!t]
\centering
\includegraphics[scale=0.41]{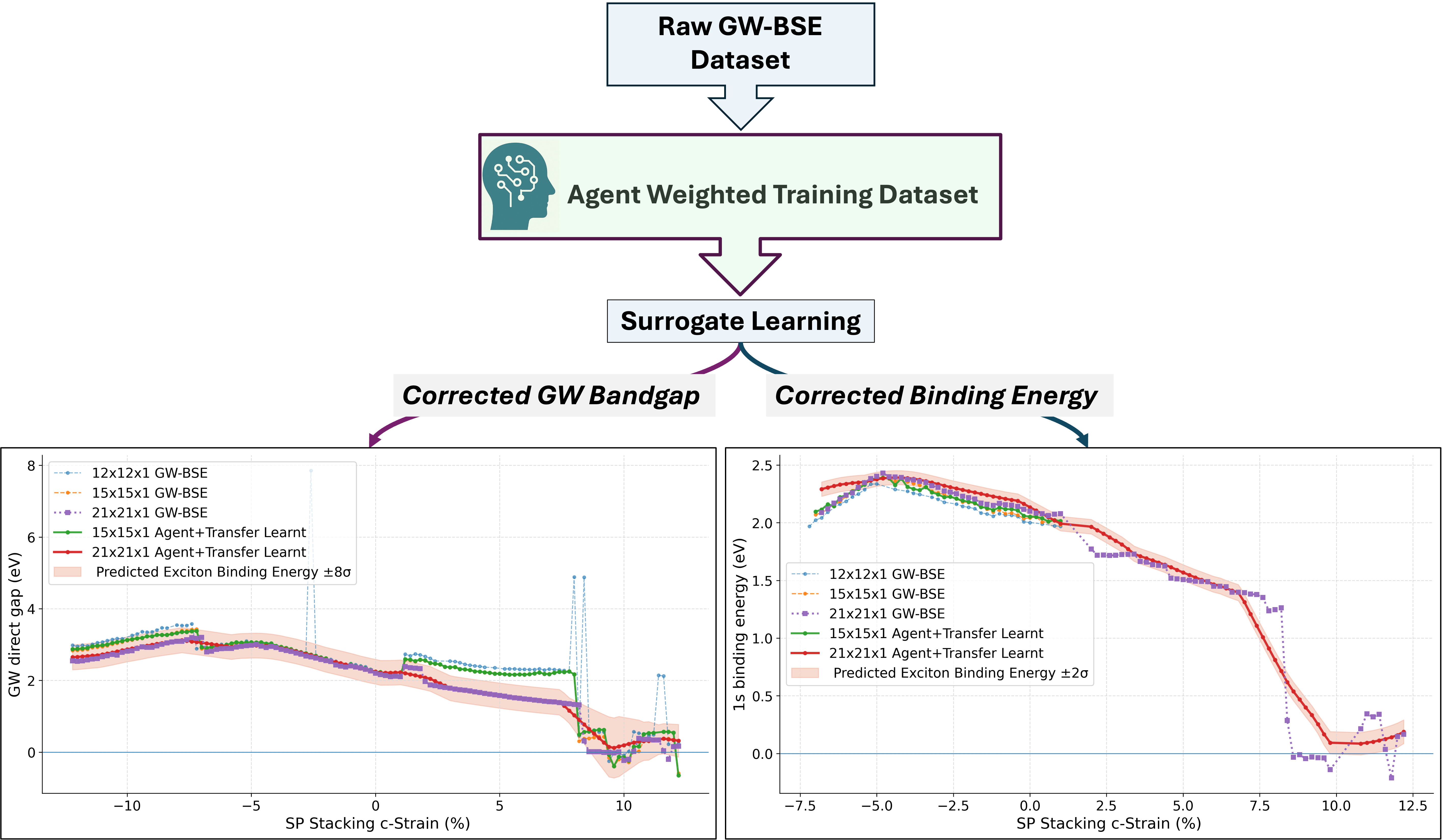}
\caption{\textbf{Closed-loop reconstruction of corrected excited-state trajectories.} Representative corrected one-dimensional reconstructions obtained after agent-guided triage, graph-based transfer learning, and Gaussian-process residual correction. The available multi-fidelity first-principles data are shown together with the final corrected trajectories and predictive uncertainty bands. The examples include corrected GW direct bandgap and corrected exciton binding energy along the SP $c$-strain branch, an environment that contains localized numerical excursions in the raw coarse-fidelity data. The corrected trajectories suppress isolated artifacts inherited from anomaly-prone inputs while preserving the physically meaningful strain dependence of the excited-state landscape. These results illustrate that sparse, well-placed higher-fidelity anchors, combined with trust-aware training and structured residual correction, can reconstruct physically plausible excited-state trends from incomplete and heterogeneous multi-fidelity data.}\label{ML_results_fig}
\end{figure*}

\begin{figure*}[!t]
\centering
\includegraphics[scale=0.55]{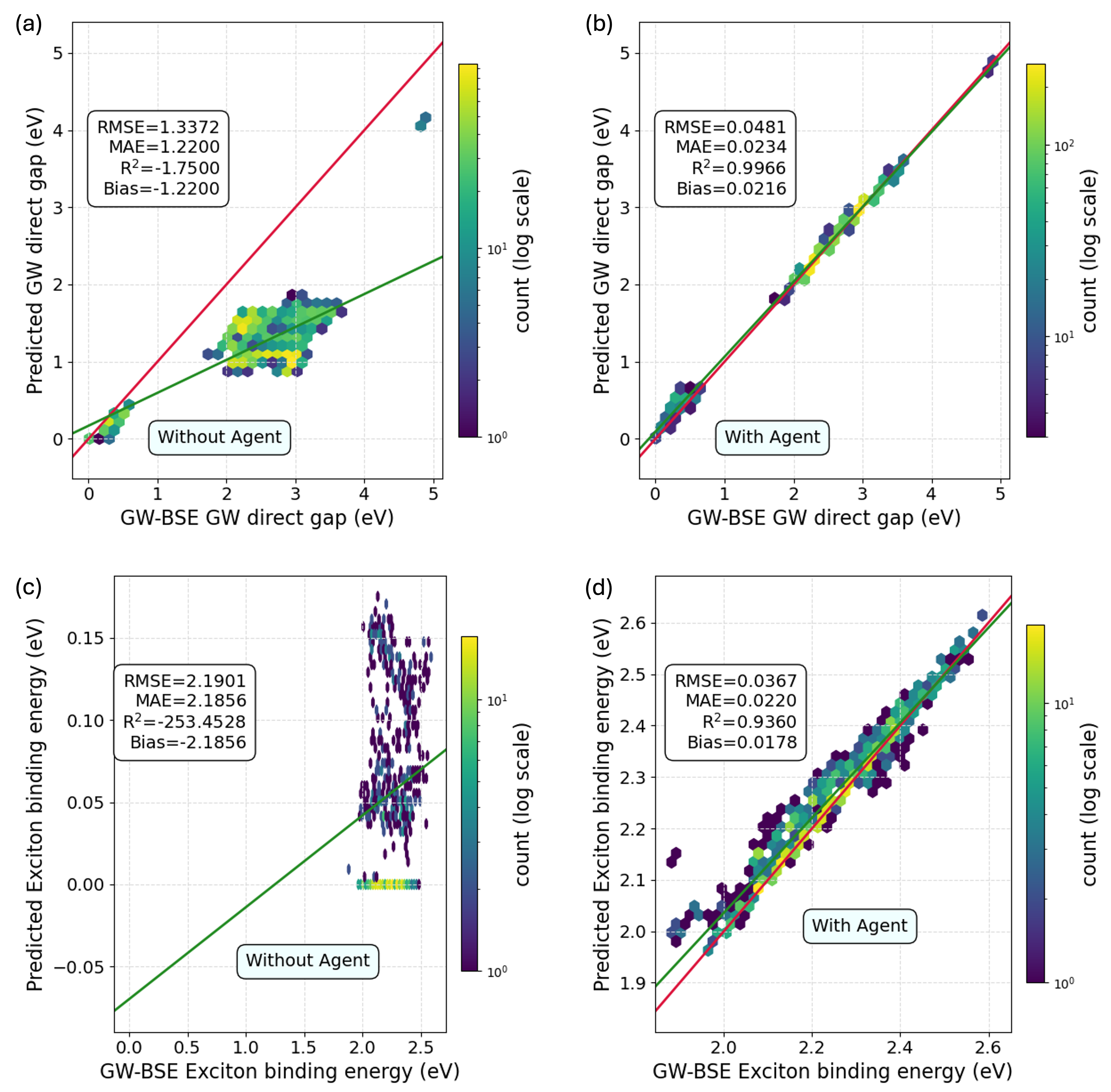}
\caption{\textbf{Agent-Guided anomaly triage improves surrogate model accuracy}
Parity plots comparing predicted and reference excited-state targets for models trained \emph{without} and \emph{with} the structured agent layer.
\textbf{(a)} GW direct bandgap without agent-guided data triage.
\textbf{(b)} GW direct bandgap with agent-guided trust weighting and anchor selection.
\textbf{(c)} Exciton binding energy without agent guidance.
\textbf{(d)} Exciton binding energy with agent guidance.
Without the agent, the surrogate is trained on a mixture of physical signal and numerically contaminated labels, leading to large systematic deviations and poor agreement with higher-fidelity references. With the agent, suspicious coarse-fidelity points are downweighted or excluded and sparse high-value anchors are prioritized, producing markedly improved agreement, reduced bias, and a substantially tighter parity distribution. The figure directly demonstrates that the agent is a necessary component for preventing the surrogate from learning numerical pathology instead of physics.}\label{Parity_comparison_fig}
\end{figure*}

\begin{figure}[h!]
\centering
\includegraphics[width=0.48\textwidth]{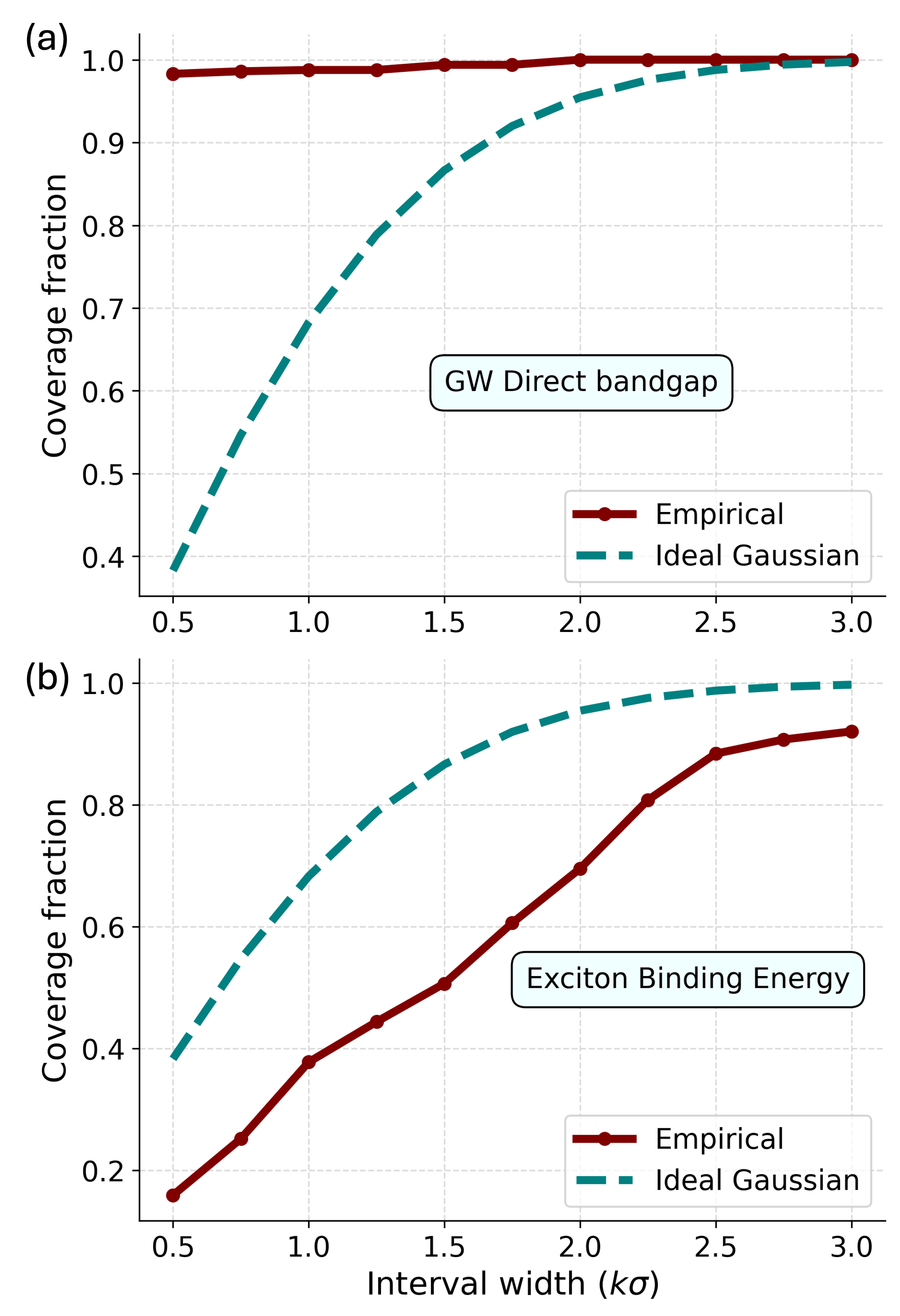}
\caption{\textbf{Uncertainty calibration of corrected excited-state predictions.} Empirical coverage curves for representative corrected targets compared with the ideal Gaussian expectation as a function of interval width $k\sigma$. (a) Corrected GW direct bandgap. (b) Corrected exciton binding energy. The empirical curves closely follow the ideal coverage over a broad range of interval widths, indicating that the predictive uncertainty is informative and reasonably calibrated rather than merely heuristic. This calibration is essential for closed-loop refinement because future anchor selection depends not only on point predictions but also on a meaningful estimate of where the corrected excited-state landscape remains uncertain.}\label{emperical_coverage_fig}
\end{figure}

\subsection{Physical Structure and Localized Numerical Failures in High-Throughput GW–BSE Calculations}

We first examine the raw excited-state landscapes produced by the stage-wise GW-BSE workflow. 
The multi-fidelity design, detailed in Methods Secs. A and B, allows the same local stacking-strain environments to be compared across reciprocal-space samplings, making it possible to separate broad strain-dependent trends from fidelity-dependent anomalies. As a result, the dataset captures a heterogeneous range of behaviors, including smooth strain trends, direct-to-indirect crossover regimes, incomplete workflow cases, and reciprocal-space convergence effects.
Figure.~\ref{DFT_results_fig} shows raw excited-state observables across the strained design space. Over broad regions, the GW direct bandgap, excitonic energy scales, and exciton binding energy evolve smoothly with strain, consistent with the expected strain sensitivity of electronic and excitonic properties in two-dimensional heterostructures \cite{farkous2020strain,yang2025exploring}.
This gradual evolution is important because it shows that the underlying excited-state space is structured and learnable. The strain dependence reflects the expected sensitivity of band alignment, dielectric screening, and electron-hole interactions to deformation. The lower-level precursor quantities in Supplementary Fig.~S1 and the band-edge evolution in Supplementary Fig.~S3 show the same broad regularity at the ground-state, direct diagonalization, and GW levels. The GW scissor corrections in Supplementary Fig.~S2 further show that the lift from lower-level electronic structure to GW is systematic, but not reducible to a single constant offset.

This mixed character is precisely the reason why in the corrected workflow, if a multi-fidelity surrogate model is trained directly from completed first-principles outputs with every completed GW-BSE calculation treated as an equally reliable training data point, the model would be asked to learn both the smooth strain-dependent physics and the localized numerical artifacts. Because these artifacts occur in structured regions rather than as homogeneous random noise, a flexible surrogate could reproduce them as part of the physical excited-state landscape. The first task of the workflow is therefore to assess the reliability of the computed labels before they are used for multi-fidelity learning.

\subsection{Breakdown of Long-Wavelength Dielectric Screening and Numerical Anomalies}

We next identify the physical and numerical origin of the most prominent anomalous regions. Figure~\ref{DFT_with_flags_fig}(a-b) focuses on SP stacking under the $c$-strain branch, where the raw GW direct bandgap develops a sharp spike and approaches a near-zero-gap regime. The exciton binding energy along the same branch shows a concurrent disruption. The simultaneous appearance of the anomaly in both quasiparticle and excitonic quantities reflects an instability that propagates through the excited-state workflow.

The corresponding dielectric-screening curves in Fig.~\ref{DFT_with_flags_fig} (c-d) provide the mechanism. In the anomaly-prone strain windows, the microscopic dielectric screening becomes strongly sensitive to reciprocal-space sampling in the small-$|G|$ regime, where $G$ denotes a reciprocal-space vector in the plane-wave representation. This behavior indicates incomplete convergence of the dielectric head. Because the GW self-energy depends directly on the screened Coulomb interaction, errors in the long-wavelength screening response can produce large errors in quasiparticle corrections. These errors then propagate into BSE-derived excitonic quantities, including the exciton binding energy. The anomaly is therefore tied to a physically interpretable numerical failure mode.

The broader screening analysis in Supplementary Fig.~S4 shows that the same mechanism applies across stacking registries, but only when the local strain configuration lies in a discontinuous region of the excited-state landscape. At the representative positive strains used for comparison, the AB and SP stackings coincide with the spike-like discontinuities shown here, and their dielectric response exhibits strong mesh-dependent variation in the small-$|G|$ regime. The AA and AA$^\prime$ stackings at the same strain values do not show the same discontinuity, and their dielectric screening is correspondingly more consistent across reciprocal-space samplings. Supplementary Figure~S4 also includes negative-strain configurations where discontinuities appear, again showing that the anomalous GW and excitonic behavior is accompanied by inconsistent long-wavelength screening. Thus, the controlling factor is whether a particular stacking-strain environment enters a numerically fragile screening regime. Irrespective of stacking registry, discontinuities in the GW gap and exciton energy are linked to inconsistencies in $\epsilon_{diag}(G)$ near the dielectric head.

This interpretation is important for the learning problem. If an anomalous point is caused by fragile dielectric convergence, then treating it as an ordinary training label is not justified. At the same time, neighboring strain points and lower-fidelity calculations may still contain useful information about the physical trend. The appropriate response is therefore to assign reliability at the level of individual points and local strain regions.

\subsection{Agent-Guided Quality Control and Data Curation}

Having established that the raw dataset contains both reliable physical trends and localized numerical failures, we next define the quality-control step used before surrogate training. We use the term agent to denote a large-language-model (LLM) based decision module, implemented as a structured GPT-5.5 API call \cite{openai2026gpt55}, that operates on precomputed numerical diagnostics of the raw GW-BSE calculations. For each flagged GW-BSE calculation, the module receives workflow-completion status, spike metrics, cross-mesh disagreement, near-zero-gap indicators, symmetry consistency, and selected numerical metadata. It returns a fixed-schema record containing a trust score, anomaly class, likely cause, recommended action, anchor priority, and training weight. These outputs determine how strongly each computed point contributes to the subsequent multi-fidelity learning step. The detailed description of developing a structured agent is in Methods secs. D and E.

The global diagnostic map in Supplementary Fig.~S12, shows the anomaly-aware spike scores and corresponding trust scores across stacking registries, strain branches, and strain values. Regions with large spike scores, strong cross-fidelity disagreement, or near-zero-gap behavior receive reduced trust, while internally consistent regions are retained as reliable evidence. Supplementary Fig.~S5 places this behavior in the context of workflow completion. The lower-level stages have broad coverage, whereas GW and BSE completion is reduced in the more difficult parts of the design space. The agent-guided layer therefore addresses two related problems at once. It limits the influence of completed but suspicious labels, and it identifies incomplete or fragile regions where higher-fidelity anchors are likely to be most valuable.

In smooth regions, dense lower-fidelity data are retained because they encode the broad strain dependence of the manifold. In anomaly-prone regions, suspicious points are downweighted, excluded, or marked for prediction rather than direct training. Points with high expected value for improving the corrected landscape are assigned higher anchor priority. This produces a curated training set that is more informative than the raw completed dataset and more selective than a purely automated pass-fail filter.

\subsection{Reconstructing Accurate Excited-State Landscapes via Multi-Fidelity Transfer Learning}

The agent-selected data are then used to reconstruct corrected excited-state landscapes. The correction model has two components. First, a graph-attention transfer model learns strain-ordered relationships across environments and reciprocal-space fidelities. It is trained in multi-fidelity delta space so that the model learns how excited-state targets change from one reciprocal-space sampling to another. This stage is described in Methods Secs. G and H. Second, a Gaussian-process residual model removes structured errors that remain after graph-based transfer and provides predictive uncertainty. This stage is described in Methods Secs. I and K.

Figure~\ref{ML_results_fig} shows representative corrected one-dimensional trajectories for the GW direct bandgap and exciton binding energy. The corrected curves are shown together with the available multi-fidelity first-principles data and predictive uncertainty bands. In each case, the corrected trajectory preserves the physically meaningful strain dependence while correcting isolated artifacts inherited from the anomaly-prone raw inputs. The correction leaves well-supported slope changes and crossover-related structure intact, while reducing the influence of points that are inconsistent with neighboring strain values, reciprocal-space trends, or agent-assigned trust.

Additional reconstructions in Supplementary Figs.~S6-S8 show that the same behavior holds for all stacking-strain environments. Across different stacking registries and strain branches, the corrected trajectories remain close to the raw data where the raw data are internally consistent, while correcting localized artifacts where the diagnostics indicate numerical fragility. The fragile cases are therefore better understood as local stacking-strain environments. In environments with stronger screening sensitivity, incomplete workflow coverage, or larger cross-fidelity disagreement, the corrected framework still reconstructs plausible GW and excitonic trends when sparse high-value anchors and residual correction are incorporated. These supplementary results show that the method operates across structurally distinct environments and across targets with different numerical difficulty.


The corrected landscape thus combines three sources of information. Dense lower-fidelity calculations define the broad strain manifold. Sparse higher-fidelity anchors constrain the most reliable high-accuracy behavior. Agent-assigned weights determine which raw labels are allowed to influence the reconstruction strongly. This combination is what makes the approach data efficient. It avoids the need for uniformly dense fine-mesh GW-BSE calculations across the entire strain grid, while also avoiding the opposite failure mode of learning directly from contaminated coarse-fidelity labels.

\subsection{Preventing Numerical Contamination via Agent-Guided Data Selection}

A direct comparison with a no-agent baseline shows why the quality-control layer is necessary. Figure~\ref{Parity_comparison_fig} compares parity plots for models trained without and with agent-guided data selection. Without the agent, all completed labels are treated too similarly, even when some of them arise from spike-like atrifacts, near-zero-gap collapse, or strong cross-mesh disagreement. The resulting parity distributions are broad and biased for both the GW direct bandgap and the exciton binding energy. This behavior shows that the surrogate is being trained on a label space that mixes physical signal with numerical contamination.

With agent-guided trust weighting and anchor selection, the parity distributions become substantially tighter. Systematic deviations are reduced and agreement with higher-fidelity references improves. The improvement comes from changing what the model is allowed to treat as reliable evidence. Suspicious data point labels are downweighted or excluded, incomplete cases are not used as direct training anchors, and higher-fidelity anchors are emphasized where they provide the most information. Thus, the agent layer changes the effective training dataset before the surrogate learns.

The trust-stratified diagnostics in Supplementary Fig.~S9 supports this interpretation. Low-trust regions have larger raw errors and require larger residual corrections, whereas high-trust regions have smaller corrections and more stable behavior. The strain-binned correction magnitudes are also nonuniform, increasing near crossover regions, fidelity-mismatch windows, and anomaly-prone strain intervals. This selectivity in the correction framework acts most strongly where the raw data are least reliable and remain conservative where the first-principles trends are well supported.

\subsection{Calibrated Excited-State Predictions for Adaptive Refinement}

Finally, we evaluate whether the corrected surrogate provides useful uncertainty estimates. This is important because the purpose of the workflow is also to support future refinement of expensive GW-BSE calculations. A model that identifies uncertain or fragile regions can guide where additional high-fidelity anchors should be placed.

Figure~\ref{emperical_coverage_fig} shows empirical coverage curves for the corrected GW direct bandgap and exciton binding energy. The observed coverage closely follows the ideal Gaussian expectation over a broad range of interval widths. This indicates that the uncertainty estimates are informative and not heuristic. The normalized residual distributions in Supplementary Fig.~S10 provide complementary evidence. After uncertainty-aware correction, the residuals are centered near zero and compressed relative to the uncorrected case, indicating reductions in both systematic bias and variance.

The uncertainty behavior is also consistent with the physical and numerical structure of the dataset. Regions with smooth strain dependence and high agent trust have smaller correction magnitudes and narrower uncertainty. Regions near direct-to-indirect crossover, workflow incompleteness, cross-mesh disagreement, or fragile screening have larger corrections and broader uncertainty. The exciton k-space spread and agent-priority trends in Supplementary Fig.~S11 show that excitonic descriptors become most informative in the same parts of the design space where the excited-state response is more sensitive. Together with the trust maps in Supplementary Fig.~S12, these results show that the corrected surrogate provides both point predictions and a meaningful estimate of where the landscape remains uncertain.

We therefore establish that the raw GW-BSE dataset contains a learnable excited-state manifold, but also localized numerical failures. The anomalies are linked to fragile long-wavelength dielectric screening. A structured agent-guided quality-control layer converts diagnostic signals into training weights and anchor priorities. Multi-fidelity transfer learning and Gaussian-process residual correction then reconstruct corrected GW bandgap and exciton binding-energy landscapes. The comparison with the no-agent baseline shows that reliability-aware data selection is necessary, and the calibration analysis shows that the corrected surrogate can support closed-loop refinement.

\section{\label{sec:discussion}DISCUSSION}

This study examines the feasibility of high-throughput GW–BSE simulations for predicting electronic properties in strained van der Waals bilayers, assisted by modern data-science methodologies. We first show that direct GW–BSE calculations are constrained not only by their substantial computational cost, but also by localized numerical artifacts that can obscure physical trends and complicate automated workflows. Most of the excited-state manifold is smooth and suitable for transfer learning, yet localized strain windows exhibit spike-like excursions, near-zero-gap behaviour and discontinuities caused by fragile long-wavelength screening. These events are structured numerical failures, not ordinary homogeneous label noise.

This distinction changes the role of multi-fidelity learning. Lower-fidelity calculations are valuable because they encode the broad strain dependence of the manifold, but their reliability is not uniform. Some regions require interpolation across fidelities, whereas others require diagnosis before the labels can be trusted. The agent-guided workflow addresses this by separating simulation, triage and prediction. First-principles calculations provide labels, anomaly diagnostics identify reliability problems, the structured agent converts those diagnostics into trust and anchor decisions, and the graph-plus-Gaussian-process model reconstructs corrected landscapes from curated evidence.

The no-agent comparison demonstrates that this triage layer is essential. A flexible surrogate trained on all points with equal trust can reproduce numerical contamination as if it were physical strain dependence. Once suspicious points are downweighted and sparse anchors are prioritized, the corrected landscapes become more accurate, physically plausible and uncertainty aware. This suggests a route toward adaptive excited-state computation in which the goal is not simply to replace expensive calculations with a surrogate, but to identify where the calculations themselves are least reliable and incorporate that diagnosis into the learning loop.

Several limitations remain. The present study focuses on one heterobilayer family, four stacking registries and two strain branches. Broader tests across other layered materials, dielectric environments and moire-related superstructures will be needed to establish generality. The framework also mitigates the consequences of numerical fragility rather than removing the underlying convergence difficulty. Improved treatments of the long-wavelength screening limit or alternative many-body starting points may reduce the problem upstream. Finally, although the agent is constrained by a fixed schema and explicit scientific rules, its recommendations depend on the quality of the anomaly signals supplied to it. Future benchmarks against deterministic triage rules and alternative acquisition strategies will help isolate the contribution of the structured reasoning layer.

Within these limits, the present workflow provides a practical strategy for reliable surrogate learning from numerically delicate excited-state calculations. Generically, it illustrates how scientific machine learning can be coupled to diagnostic information from the data-generation process itself, allowing simulation, diagnosis, learning and targeted refinement to operate as a single adaptive loop.
The value of the present results is that they turn high-throughput GW-BSE calculations from a collection of expensive and unevenly reliable outputs into a reliability-aware excited-state dataset that can support predictive learning. The framework identifies which completed calculations provide trustworthy training evidence, limits the influence of numerically fragile data points, and reconstructs corrected GW bandgap and exciton-binding landscapes with calibrated uncertainty. This makes the approach useful not only for the strained MoS$_2$-WS$_2$ bilayers studied here, but also as a template for other materials problems in which expensive first-principles calculations fail through subtle convergence errors rather than obvious job failure. In the future works, the same strategy can be extended to broader families of layered semiconductors, dielectric environments, moire heterostructures, defects, and alloyed systems. More generally, it points toward an adaptive mode of excited-state materials discovery in which simulation, diagnostic evaluation, agent-guided data selection, surrogate learning, and targeted high-fidelity refinement form a closed loop for building reliable large-scale datasets.

\section{Methods}\label{methods}

\subsection{High-throughput multi-fidelity first-principles workflow}

We constructed a multi-fidelity excited-state dataset for strained MoS$_2$-WS$_2$ bilayers by combining first-principles calculations across all the stacking-strain-fidelity environments. For each strained configuration, the workflow proceeds through four stages: ground-state density functional theory, exact diagonalization of the one-electron manifold, GW quasiparticle correction, and Bethe--Salpeter equation analysis of optical excitations. Reciprocal-space sampling is varied over $12\times12\times1$, $15\times15\times1$, and $21\times21\times1$ in order to expose and reduce $k$-mesh-dependent anomalies in the excited-state response. This workflow is the basis for the multi-fidelity dataset discussed in the Results section and summarized schematically in Fig.~\ref{Schematic_fig}. 

For each stacking registry, we generated two families of strained structures over $\varepsilon\in[-12\%,12\%]$. The first branch corresponds to in-plane biaxial strain, denoted $ab$ strain, in which the in-plane lattice vectors were scaled uniformly. The second branch corresponds to a $c$-strain protocol designed to modulate the interlayer separation through in-plane counter-scaling followed by constrained relaxation. Each stacking and strain-branch pair defines a strain-ordered environment. The resulting dataset spans smooth regimes, crossover regions, and anomaly-prone windows of the strained excited-state manifold. Supporting Figs.~S1--S5 summarize the lower-level precursor trends, stage attrition, and dielectric-convergence sensitivity across this design space.

\subsection{Ground-state, exact-diagonalization, GW, and BSE calculations}

All first-principles calculations were performed using the Vienna \textit{Ab initio} Simulation Package \cite{hafner2008ab} with spin--orbit coupling retained throughout the workflow. Ground-state calculations were used to relax the strained structures and generate the charge densities and wavefunctions required for the later stages. The ground-state production settings used a plane-wave cutoff of 260~eV, Gaussian smearing of width 0.05~eV, an electronic convergence threshold of $10^{-6}$~eV, and ionic relaxation until the maximum force was below 0.02~eV~\AA$^{-1}$. Wavefunctions and charge densities were written to disk and reused in subsequent stages. Complete stage-specific settings are provided in Supplementary Methods Sec.~1.
The exact-diagonalization stage was performed on top of the converged ground-state charge density using a fixed-density one-electron calculation with an enlarged empty-state manifold. This stage provides the precursor one-electron structure together with dielectric and optical descriptors used downstream in the workflow. Quasiparticle corrections were then computed within an eigenvalue-self-consistent GW scheme. Because the strained dielectric response is especially sensitive in the long-wavelength limit, incomplete convergence in this regime can induce large excursions in the screened interaction and the resulting quasiparticle gaps, as illustrated in Fig.~\ref{DFT_with_flags_fig} and Fig.~S4. 

Excitonic properties were obtained by solving the Bethe--Salpeter equation on top of the GW-corrected one-electron manifold. From the resulting spectra we extracted the lowest bright exciton energy $E_1$, its oscillator strength, and the direct exciton binding energy
\begin{equation}
E_b = E^{\mathrm{GW}}_{g,\mathrm{dir}} - E_1.
\label{eq:binding_energy}
\end{equation}
In addition, we retained reciprocal-space descriptors derived from the excitonic amplitudes, including the exciton $k$-space centroid and $k$-space spread, which were later used in anomaly detection and residual correction. Supporting Fig.~S11 and Appendix Fig.~A2 provide further information on excitonic descriptors and optical-brightness trends across the strained design space.

\subsection{Target definitions and training eligibility}

The implementation defines three supervised targets,
\begin{equation}
Y \in \left\{ E^{\mathrm{GW}}_{g,\mathrm{dir}},\; E^{\mathrm{GW}}_{g,\mathrm{fund}},\; E^{1s}_{b} \right\}.
\label{eq:targets}
\end{equation}
For the direct and fundamental GW gaps, the learning target is represented as a correction relative to the exact-diagonalization reference,
\begin{equation}
y_{\mathrm{model}} = y_{\mathrm{GW}} - y_{\mathrm{KS,exact}},
\label{eq:delta_target}
\end{equation}
whereas the exciton binding energy is modeled directly in energy units. Near-zero-gap thresholds are used only for anomaly detection and trust scoring, not for redefining the target itself. This distinction is important for the Results section, where near-zero-gap behavior is interpreted as a sign of numerical fragility rather than as a target transformation. 

Training eligibility is stage dependent. GW targets require successful completion of the exact-diagonalization and GW stages together with finite target values. The binding-energy target additionally requires completion of the BSE stage. Points with incomplete stage execution can still enter the workflow as prediction-only cases, but they are not admitted as direct training anchors. These eligibility rules are part of the scientific triage process described below and help determine which labels are allowed to influence the corrected surrogate.

\subsection{Node representation and anomaly-oriented feature construction}

Each strained calculation is represented as a node indexed by stacking, strain branch, strain value, stage completion, and reciprocal-space fidelity. Node features combine band-edge descriptors, quasiparticle gaps, excitonic observables, reciprocal-space descriptors, strain metadata, and selected numerical settings parsed from the stage-specific inputs. Band-edge features include direct and indirect gaps from the ground-state, exact-diagonalization, and GW stages together with derived scissor-like corrections. Excitonic features include the lowest bright exciton energy, higher excitonic levels when available, oscillator strengths, and reciprocal-space descriptors such as $k$ centroid and $k$ spread. Numerical settings are retained as auxiliary variables rather than as physical targets. 

To expose fidelity mismatch explicitly, we construct cross-mesh engineered features from the node table:
\begin{align}
\Delta_{12\rightarrow 15} &= Y_{15} - Y_{12}, \\
\Delta_{15\rightarrow 21} &= Y_{21} - Y_{15}, \\
\Delta_{12\rightarrow 21} &= Y_{21} - Y_{12},
\end{align}
together with the cross-mesh standard deviation
\begin{equation}
\sigma_{\mathrm{across}\ k} = \mathrm{std}(Y_{12},Y_{15},Y_{21}).
\end{equation}
A symmetry-based diagnostic is also formed by comparing each point with its sign-inverted strain partner within the same environment. This enters the workflow only as a soft anomaly signal rather than as a hard physical constraint. These engineered features define the quantitative basis for the anomaly diagnostics discussed in the Results section. 

\subsection{Spike diagnostics and fragile-regime flags}

To quantify abrupt local excursions, the workflow computes an environment-level spike score from the normalized second finite difference along the strain axis. For a strain-ordered sequence $y(\varepsilon_i)$, the discrete second difference is
\begin{equation}
\Delta^2 y_i = y_{i+1} - 2y_i + y_{i-1}.
\end{equation}
The spike score is then defined as
\begin{equation}
S_{\mathrm{spike}} =
\frac{\max_i |\Delta^2 y_i|}
{\mathrm{median}_i(|\Delta^2 y_i|) + 10^{-12}}.
\label{eq:spike_score}
\end{equation}
If the number of valid points in a sequence is below six, the score is set to zero. This diagnostic is one of the principal signals used to identify the anomaly-prone regions shown in Fig.~\ref{DFT_with_flags_fig} and mapped globally in Fig.~S12. 

Points are sent to the agent if they belong to environments with large spike scores, near-zero-gap behavior, large cross-mesh disagreement, missing target completion, large symmetry discrepancies, band-count mismatch across $k$ meshes, reduced GW band count, or explicitly flagged extreme-strain regimes. An additional untrusted-zero-gap candidate flag is used for extreme-strain points with near-zero gaps when they coincide with high spike score or strong cross-mesh inconsistency. These cases are not treated as ordinary training anchors in the corrected-target workflow. In this way, the anomaly-detection stage operationalizes the distinction drawn in the Results between physically meaningful structure and localized numerical pathology. 

\subsection{Structured agent for anomaly-aware scientific triage}

We introduce a structured agent layer that is called only on points satisfying anomaly-gating criteria. The agent is given stage completion, spike diagnostics, cross-fidelity disagreement, near-zero-gap flags, symmetry consistency, and selected numerical-setting information. Under a fixed schema, it returns trust scores, anomaly classes, likely causes, recommended actions, anchor priorities, training weights, and free-text notes. The allowed anomaly classes include ordinary points, spike or jump behavior, zero-gap regime, cross-mesh disagreement, fidelity mismatch, incomplete workflow, likely physical transition, and unknown cases. The likely-cause field distinguishes small-$G$ screening sensitivity, $k$-mesh sampling, reduced corrected bands, stage incompleteness, possible band crossover, parser issues, and unknown causes. The full schema and representative prompt are provided in Appendix~B. 

The scientific instructions supplied to the agent encode the main failure modes relevant to strained GW--BSE data. These include the facts that GW jumps can arise from dielectric-head instability near $G\rightarrow 0$, that denser $k$ meshes may reduce but not eliminate such anomalies, that reduced-band-count behavior should be treated as a fidelity mismatch, and that isolated near-zero-gap collapse in extreme-strain regimes should not be overtrusted. Symmetry is used only as a soft check, and the GW strain trend is expected to broadly follow the lower-level trend in smooth and well-converged regions, although not necessarily as a constant scissor shift. These rules are the basis for the scientific-triage role highlighted in the Results section. 

Crucially, the agent is restricted to data quality control and anchor selection. Its recommended actions include retaining points for training, downweighting suspicious points, excluding pathological cases, marking incomplete cases as prediction-only, and requesting higher-fidelity anchor calculations. This restricted role is scientifically important. It keeps first-principles simulation, anomaly diagnosis, and numerical prediction conceptually separate, which is exactly the interpretation emphasized in the Results and Discussion. The corresponding global spike and trust maps are shown in Fig.~S12, while trust-stratified error behavior is shown in Fig.~S9. 

\subsection{Training weights and anchor handling}

The agent returns both a trust score and a training weight. These are combined to form the effective fitting weight for the corrected baseline. If a point is flagged as a spike, a near-zero-gap point, a large jump, or an untrusted zero-gap candidate, its fitting weight is strongly reduced. Suspicious points can therefore remain visible to the model while their influence on the corrected fit is limited. This weighting strategy is central to the agent-versus-no-agent comparison presented in Fig.~\ref{Parity_comparison_fig}. 

The workflow constructs a corrected intermediate-fidelity baseline from the $15\times15\times1$ data and uses the learned deltas to infer corrected $12\times12\times1$ and corrected $21\times21\times1$ trajectories. This design makes the intermediate $k$ mesh a bridge between the coarse and fine fidelities and prevents the final corrected targets from depending on any single raw anomalous curve. Sparse higher-fidelity anchors are therefore used only where they add the most value, rather than being distributed uniformly across the entire strain grid. This is the anchor-handling logic referred to in the Results section when discussing data efficiency and targeted refinement.

\subsection{Graph-attention transfer model}

Each environment, defined by a stacking and strain-branch pair, is modeled as a strain-ordered sequence of nodes. The baseline predictor is a graph-attention model that maps the node feature vector to a mean prediction $\hat{y}_{\mathrm{GAT}}$ for each target. The model uses projected feature embeddings, positional encoding along the strain axis, and attention-based message passing across the sequence. Environment-level splits are used in training and validation so that complete environments, rather than individual strain points, are assigned to train and validation groups. This design avoids leakage across strain trajectories and tests generalization across environments rather than interpolation within a single curve. 

In the multi-fidelity setting, the graph model is trained in delta space. For two reciprocal-space samplings $k_a$ and $k_b$, we define the inter-fidelity correction
\begin{equation}
\Delta_{k_a\rightarrow k_b} = Y_{k_b} - Y_{k_a}.
\label{eq:delta_space}
\end{equation}
The graph model therefore learns how to transfer from lower-fidelity to higher-fidelity landscapes rather than fitting the highest-fidelity target from scratch. Lower-fidelity calculations provide dense strain coverage, whereas higher-fidelity GW and BSE calculations provide sparse but more reliable anchors. This is the transfer-learning stage referred to in the Results section when discussing reconstruction of corrected strain trends in Fig.~\ref{ML_results_fig} and Figs.~S6-S8. 

\subsection{Causal-structure-informed Gaussian-process residual correction}

To remove the structured error that remains after graph inference, we fit a Gaussian-process residual model on a compact set of physically meaningful parent variables selected through causal-structure analysis. Let $x_{\mathrm{Pa}(Y)}$ denote the parent-variable set for target $Y$. The corrected predictor is written as
\begin{equation}
\hat{Y} = \hat{Y}_{\mathrm{GAT}} + r\!\left(x_{\mathrm{Pa}(Y)}\right),
\label{eq:gp_corrected}
\end{equation}
where $r$ denotes the Gaussian-process residual model. For each target, the residual is defined as
\begin{equation}
r = Y - \hat{Y}_{\mathrm{GAT}}.
\label{eq:residual}
\end{equation}

Constraint-based causal discovery is performed on a compact set of physically meaningful variables, including GW gaps, lower-fidelity gaps, scissor terms, oscillator strengths, and reciprocal-space exciton descriptors. The residual correction is then restricted to the selected parent set rather than being fitted over the full feature bank. This restriction reduces the risk that the final correction drifts toward nuisance numerical settings and instead keeps the correction more interpretable and transferable. Gaussian-process regression is carried out using kernels built from a constant term, a radial-basis component, and a white-noise term, with standardization applied to the parent features before fitting. The workflow also applies one-dimensional Gaussian-process correction along the strain axis to produce corrected trajectories and predictive uncertainty bands. This is the residual-correction stage referred to in the Results section when describing selective correction magnitude, the reconstructions in Fig.~\ref{ML_results_fig}, and the strain-binned diagnostics in Fig.~S9. 

\subsection{Soft guardrail against unphysical near-zero GW gaps}

For GW targets only, the workflow includes a soft guardrail designed to prevent unphysical collapse of the corrected gap in fragile extreme-strain regimes. If a predicted fine-fidelity GW gap satisfies all of the following conditions,
\begin{enumerate}
\item the strain magnitude is large,
\item the predicted gap is near zero,
\item the point has been flagged as fragile,
\end{enumerate}
then the prediction is raised to a soft floor defined from local neighboring values and the corresponding exact-diagonalization reference. The predictive uncertainty is also enlarged at that point. 

This step is intended to suppress clearly unphysical numerical collapse rather than distort legitimate physical transitions. In the main text, near-zero-gap regimes are therefore interpreted conservatively. They are preserved when supported by cross-fidelity consistency and surrounding evidence, but are prevented from dominating the corrected landscape when the anomaly diagnostics indicate numerical fragility. That interpretation is consistent with the anomaly examples in Fig.~\ref{DFT_with_flags_fig}. 

\subsection{Corrected targets, evaluation protocol, and uncertainty calibration}

The principal corrected targets in this work are the GW direct bandgap and the direct exciton binding energy. For each stacking, strain branch, strain value, and current fidelity, the final corrected estimate combines the available first-principles value at that fidelity, the graph-based transfer prediction across the multi-fidelity manifold, and the Gaussian-process residual correction conditioned on the selected parent variables. This yields a corrected excited-state landscape that is more robust to isolated spike artifacts, stage incompleteness, and fidelity mismatch than any single raw $k$-mesh dataset alone. 

Model quality is evaluated against the highest-fidelity available references using root mean square error, mean absolute error, coefficient of determination, and bias. We further examine one-dimensional strain reconstructions, parity behavior, normalized residual distributions, empirical uncertainty coverage, and trust-stratified error summaries to verify that the correction improves anomaly-prone regions without distorting well-behaved parts of the strain manifold. These are the evaluation procedures referred to in the Results section when discussing Figs.~4-6, S9, and S10. 

Empirical uncertainty calibration is assessed by comparing the observed coverage of predictive intervals against the ideal Gaussian expectation over a range of interval widths. The normalized residuals are also analyzed in the form $(\hat{y}-y)/\sigma$ in order to determine whether the predictive uncertainty is informative rather than merely heuristic. These diagnostics support the uncertainty claims made in the Results section and are summarized in Fig.~\ref{emperical_coverage_fig} and Fig.~S10. 



\section{Declarations}

\subsection{Data Availability}
The representative datasets associated with this study are currently undergoing institutional review. They will be released through a public repository upon completion of the required approval process.

\subsection{Code Availability}
The code associated with this work is currently undergoing institutional review. It will be released through a public repository after completion of the required institutional approval process.

\subsection{Acknowledgments}

This work at Los Alamos was carried out under the auspices of the U.S. Department of Energy (DOE) National Nuclear Security Administration (NNSA) under Contract No. 89233218CNA000001. It was supported by Center for Integrated Nanotechnologies (CINT), a DOE BES user facility, in partnership with the LANL Institutional Computing Program for computational resources. C.A.L. is supported by LANL LDRD program.

\subsection{Author Contribution}

A.N. developed the codes for the high-throughput multi-fidelity workflow, the agent-guided framework, and the transfer-learning models, and drafted the manuscript. A.F and C.A.L contributed to scientific discussions and edited the manuscript. J.X.Z. and S.T. designed and supervised the project, contributed to scientific discussions, and edited the manuscript.

\subsection{Conflict of interest/Competing interests}

The authors declare no competing interests.

\onecolumngrid
\noindent\begin{center}
\textbf{\large References}
\end{center}
\twocolumngrid
\bibliography{apssamp}

@article{hafner2008ab,
  title={Ab-initio simulations of materials using VASP: Density-functional theory and beyond},
  author={Hafner, J{\"u}rgen},
  journal={Journal of computational chemistry},
  volume={29},
  number={13},
  pages={2044--2078},
  year={2008},
  publisher={Wiley Online Library}
}

@article{cutkosky1954solutions,
  title={Solutions of a Bethe-Salpeter equation},
  author={Cutkosky, RE66966},
  journal={Physical Review},
  volume={96},
  number={4},
  pages={1135},
  year={1954},
  publisher={APS}
}

@article{velivckovic2017graph,
  title={Graph attention networks},
  author={Veli{\v{c}}kovi{\'c}, Petar and Cucurull, Guillem and Casanova, Arantxa and Romero, Adriana and Lio, Pietro and Bengio, Yoshua},
  journal={arXiv preprint arXiv:1710.10903},
  year={2017}
}

@article{deringer2021gaussian,
  title={Gaussian process regression for materials and molecules},
  author={Deringer, Volker L and Bart{\'o}k, Albert P and Bernstein, Noam and Wilkins, David M and Ceriotti, Michele and Cs{\'a}nyi, G{\'a}bor},
  journal={Chemical reviews},
  volume={121},
  number={16},
  pages={10073--10141},
  year={2021},
  publisher={ACS Publications}
}

@article{williams1995gaussian,
  title={Gaussian processes for regression},
  author={Williams, Christopher and Rasmussen, Carl},
  journal={Advances in neural information processing systems},
  volume={8},
  year={1995}
}

@article{manzeli20172d,
  title={2D transition metal dichalcogenides},
  author={Manzeli, Sajedeh and Ovchinnikov, Dmitry and Pasquier, Diego and Yazyev, Oleg V and Kis, Andras},
  journal={Nature Reviews Materials},
  volume={2},
  number={8},
  pages={1--15},
  year={2017},
  publisher={Nature Publishing Group}
}

@article{mak2016photonics,
  title={Photonics and optoelectronics of 2D semiconductor transition metal dichalcogenides},
  author={Mak, Kin Fai and Shan, Jie},
  journal={Nature Photonics},
  volume={10},
  number={4},
  pages={216--226},
  year={2016},
  publisher={Nature Publishing Group UK London}
}

@article{ye2014probing,
  title={Probing excitonic dark states in single-layer tungsten disulphide},
  author={Ye, Ziliang and Cao, Ting and O’brien, Kevin and Zhu, Hanyu and Yin, Xiaobo and Wang, Yuan and Louie, Steven G and Zhang, Xiang},
  journal={Nature},
  volume={513},
  number={7517},
  pages={214--218},
  year={2014},
  publisher={Nature Publishing Group UK London}
}

@article{chaves2020bandgap,
  title={Bandgap engineering of two-dimensional semiconductor materials},
  author={Chaves, A and Azadani, Javad G and Alsalman, Hussain and Da Costa, DR and Frisenda, R and Chaves, AJ and Song, Seung Hyun and Kim, Young Duck and He, Daowei and Zhou, Jiadong and others},
  journal={npj 2D Materials and Applications},
  volume={4},
  number={1},
  pages={29},
  year={2020},
  publisher={Nature Publishing Group UK London}
}

@article{peng2020strain,
  title={Strain engineering of 2D semiconductors and graphene: from strain fields to band-structure tuning and photonic applications},
  author={Peng, Zhiwei and Chen, Xiaolin and Fan, Yulong and Srolovitz, David J and Lei, Dangyuan},
  journal={Light: Science \& Applications},
  volume={9},
  number={1},
  pages={190},
  year={2020},
  publisher={Nature Publishing Group UK London}
}

@article{westermayr2020machine,
  title={Machine learning for electronically excited states of molecules},
  author={Westermayr, Julia and Marquetand, Philipp},
  journal={Chemical Reviews},
  volume={121},
  number={16},
  pages={9873--9926},
  year={2020},
  publisher={ACS Publications}
}

@article{ahn2017strain,
  title={Strain-engineered growth of two-dimensional materials},
  author={Ahn, Geun Ho and Amani, Matin and Rasool, Haider and Lien, Der-Hsien and Mastandrea, James P and Ager III, Joel W and Dubey, Madan and Chrzan, Daryl C and Minor, Andrew M and Javey, Ali},
  journal={Nature communications},
  volume={8},
  number={1},
  pages={608},
  year={2017},
  publisher={Nature Publishing Group UK London}
}

@article{chen2020chiral,
  title={Chiral coupling of valley excitons and light through photonic spin--orbit interactions},
  author={Chen, Peigang and Lo, Tsz Wing and Fan, Yulong and Wang, Shubo and Huang, Haitao and Lei, Dangyuan},
  journal={Advanced Optical Materials},
  volume={8},
  number={5},
  pages={1901233},
  year={2020},
  publisher={Wiley Online Library}
}

@article{withers2015light,
  title={Light-emitting diodes by band-structure engineering in van der Waals heterostructures},
  author={Withers, Freddie and Del Pozo-Zamudio, Osvaldo and Mishchenko, Artem and Rooney, Aiden P and Gholinia, Ali and Watanabe, K and Taniguchi, Takashi and Haigh, Sarah J and Geim, AK and Tartakovskii, AI and others},
  journal={Nature materials},
  volume={14},
  number={3},
  pages={301--306},
  year={2015},
  publisher={Nature Publishing Group UK London}
}

@article{qi2023recent,
  title={Recent progress in strain engineering on van der Waals 2D materials: Tunable electrical, electrochemical, magnetic, and optical properties},
  author={Qi, Yaping and Sadi, Mohammad A and Hu, Dan and Zheng, Ming and Wu, Zhenping and Jiang, Yucheng and Chen, Yong P},
  journal={Advanced Materials},
  volume={35},
  number={12},
  pages={2205714},
  year={2023},
  publisher={Wiley Online Library}
}

@article{du2021strain,
  title={Strain engineering in 2D material-based flexible optoelectronics},
  author={Du, Junli and Yu, Huihui and Liu, Baishan and Hong, Mengyu and Liao, Qingliang and Zhang, Zheng and Zhang, Yue},
  journal={Small Methods},
  volume={5},
  number={1},
  pages={2000919},
  year={2021},
  publisher={Wiley Online Library}
}

@article{leng2016gw,
  title={GW method and Bethe--Salpeter equation for calculating electronic excitations},
  author={Leng, Xia and Jin, Fan and Wei, Min and Ma, Yuchen},
  journal={Wiley Interdisciplinary Reviews: Computational Molecular Science},
  volume={6},
  number={5},
  pages={532--550},
  year={2016},
  publisher={Wiley Online Library}
}

@article{biswas2023py,
  title={py GWBSE: a high throughput workflow package for GW-BSE calculations},
  author={Biswas, Tathagata and Singh, Arunima K},
  journal={npj Computational Materials},
  volume={9},
  number={1},
  pages={22},
  year={2023},
  publisher={Nature Publishing Group UK London}
}

@article{he2016strain,
  title={Strain-induced electronic structure changes in stacked van der Waals heterostructures},
  author={He, Yongmin and Yang, Yang and Zhang, Zhuhua and Gong, Yongji and Zhou, Wu and Hu, Zhili and Ye, Gonglan and Zhang, Xiang and Bianco, Elisabeth and Lei, Sidong and others},
  journal={Nano letters},
  volume={16},
  number={5},
  pages={3314--3320},
  year={2016},
  publisher={ACS Publications}
}

@article{varrassi2025automated,
  title={Automated workflow for accurate high-throughput GW calculations using plane waves},
  author={Varrassi, Lorenzo and Ellinger, Florian and Flage-Larsen, Espen and Wolloch, Michael and Kresse, Georg and Marzari, Nicola and Franchini, Cesare},
  journal={npj Computational Materials},
  volume={11},
  number={1},
  pages={351},
  year={2025},
  publisher={Nature Publishing Group UK London}
}

@article{baum2026transfer,
  title={Transfer learning of GW Bethe--Salpeter equation excitation energies},
  author={Baum, Dario and F{\"o}rster, Arno and Visscher, Lucas},
  journal={Chemical Science},
  year={2026},
  publisher={Royal Society of Chemistry}
}

@article{vinod2023multifidelity,
  title={Multifidelity machine learning for molecular excitation energies},
  author={Vinod, Vivin and Maity, Sayan and Zaspel, Peter and Kleinekath{\"o}fer, Ulrich},
  journal={Journal of Chemical Theory and Computation},
  volume={19},
  number={21},
  pages={7658--7670},
  year={2023},
  publisher={ACS Publications}
}

@article{yang2022multi,
  title={Multi-fidelity machine learning models for structure--property mapping of organic electronics},
  author={Yang, Chih-Hsuan and Pokuri, Balaji Sesha Sarath and Lee, Xian Yeow and Balakrishnan, Sangeeth and Hegde, Chinmay and Sarkar, Soumik and Ganapathysubramanian, Baskar},
  journal={Computational Materials Science},
  volume={213},
  pages={111599},
  year={2022},
  publisher={Elsevier}
}

@article{pak2017strain,
  title={Strain-mediated interlayer coupling effects on the excitonic behaviors in an epitaxially grown MoS2/WS2 van der Waals heterobilayer},
  author={Pak, Sangyeon and Lee, Juwon and Lee, Young-Woo and Jang, A-Rang and Ahn, Seongjoon and Ma, Kyung Yeol and Cho, Yuljae and Hong, John and Lee, Sanghyo and Jeong, Hu Young and others},
  journal={Nano letters},
  volume={17},
  number={9},
  pages={5634--5640},
  year={2017},
  publisher={ACS Publications}
}

@article{aryasetiawan1998gw,
  title={The GW method},
  author={Aryasetiawan, Ferdi and Gunnarsson, Olle},
  journal={Reports on progress in Physics},
  volume={61},
  number={3},
  pages={237--312},
  year={1998}
}

@article{gan20222d,
  title={2D materials-enabled optical modulators: From visible to terahertz spectral range},
  author={Gan, Xuetao and Englund, Dirk and Van Thourhout, Dries and Zhao, Jianlin},
  journal={Applied Physics Reviews},
  volume={9},
  number={2},
  year={2022},
  publisher={AIP Publishing}
}

@article{gopalan2022theoretical,
  title={Theoretical study of electronic transport in two-dimensional transition metal dichalcogenides: Effects of the dielectric environment},
  author={Gopalan, Sanjay and Van de Put, Maarten L and Gaddemane, Gautam and Fischetti, Massimo V},
  journal={Physical Review Applied},
  volume={18},
  number={5},
  pages={054062},
  year={2022},
  publisher={APS}
}

@article{zhang2023effect,
  title={Effect of dynamical screening in the Bethe-Salpeter framework: Excitons in crystalline naphthalene},
  author={Zhang, Xiao and Leveillee, Joshua A and Schleife, Andr{\'e}},
  journal={Physical Review B},
  volume={107},
  number={23},
  pages={235205},
  year={2023},
  publisher={APS}
}

@article{gulans2014exciting,
  title={Exciting: a full-potential all-electron package implementing density-functional theory and many-body perturbation theory},
  author={Gulans, Andris and Kontur, Stefan and Meisenbichler, Christian and Nabok, Dmitrii and Pavone, Pasquale and Rigamonti, Santiago and Sagmeister, Stephan and Werner, Ute and Draxl, Claudia},
  journal={Journal of Physics: Condensed Matter},
  volume={26},
  number={36},
  pages={363202},
  year={2014},
  publisher={IOP Publishing}
}

@article{druppel2018electronic,
  title={Electronic excitations in transition metal dichalcogenide monolayers from an LDA+ GdW approach},
  author={Dr{\"u}ppel, Matthias and Deilmann, Thorsten and Noky, Jonathan and Marauhn, Philipp and Kr{\"u}ger, Peter and Rohlfing, Michael},
  journal={Physical Review B},
  volume={98},
  number={15},
  pages={155433},
  year={2018},
  publisher={APS}
}

@article{louis2020graph,
  title={Graph convolutional neural networks with global attention for improved materials property prediction},
  author={Louis, Steph-Yves and Zhao, Yong and Nasiri, Alireza and Wang, Xiran and Song, Yuqi and Liu, Fei and Hu, Jianjun},
  journal={Physical Chemistry Chemical Physics},
  volume={22},
  number={32},
  pages={18141--18148},
  year={2020},
  publisher={Royal Society of Chemistry}
}

@inproceedings{wang2019heterogeneous,
  title={Heterogeneous graph attention network},
  author={Wang, Xiao and Ji, Houye and Shi, Chuan and Wang, Bai and Ye, Yanfang and Cui, Peng and Yu, Philip S},
  booktitle={The world wide web conference},
  pages={2022--2032},
  year={2019}
}

@article{heinze2018causal,
  title={Causal structure learning},
  author={Heinze-Deml, Christina and Maathuis, Marloes H and Meinshausen, Nicolai},
  journal={Annual Review of Statistics and Its Application},
  volume={5},
  pages={371--391},
  year={2018},
  publisher={Annual Reviews}
}

@article{hu2022causality,
  title={Causality-driven hierarchical structure discovery for reinforcement learning},
  author={Hu, Xing and Zhang, Rui and Tang, Ke and Guo, Jiaming and Yi, Qi and Chen, Ruizhi and Du, Zidong and Li, Ling and Guo, Qi and Chen, Yunji and others},
  journal={Advances in Neural Information Processing Systems},
  volume={35},
  pages={20064--20076},
  year={2022}
}

@misc{openai2026gpt55,
  title = {GPT-5.5 (Large Language Model)},
  author = {{OpenAI}},
  year = {2026},
  url = {https://openai.com},
  note = {Accessed: 2026-05-25}
}

@article{hedin1965new,
  title={New method for calculating the one-particle Green's function with application to the electron-gas problem},
  author={Hedin, Lars},
  journal={Physical Review},
  volume={139},
  number={3A},
  pages={A796},
  year={1965},
  publisher={APS}
}

@article{aulbur1999quasiparticle,
  title={Quasiparticle calculations in solids},
  author={Aulbur, Wilfried G and J{\"o}nsson, Lars and Wilkins, John W},
  journal={Solid State Physics},
  volume={54},
  pages={1--218},
  year={1999}
}

@article{kirklin2015open,
  title={The Open Quantum Materials Database (OQMD): assessing the accuracy of DFT formation energies},
  author={Kirklin, Scott and Saal, James E and Meredig, Bryce and Thompson, Alex and Doak, Jeff W and Aykol, Muratahan and R{\"u}hl, Stephan and Wolverton, Chris},
  journal={npj Computational Materials},
  volume={1},
  number={1},
  pages={1--15},
  year={2015},
  publisher={Nature Publishing Group}
}

@article{horton2019high,
  title={High-throughput prediction of the ground-state collinear magnetic order of inorganic materials using density functional theory},
  author={Horton, Matthew Kristofer and Montoya, Joseph Harold and Liu, Miao and Persson, Kristin Aslaug},
  journal={npj Computational Materials},
  volume={5},
  number={1},
  pages={64},
  year={2019},
  publisher={Nature Publishing Group UK London}
}

@article{jain2013commentary,
  title={Commentary: The Materials Project: A materials genome approach to accelerating materials innovation},
  author={Jain, Anubhav and Ong, Shyue Ping and Hautier, Geoffroy and Chen, Wei and Richards, William Davidson and Dacek, Stephen and Cholia, Shreyas and Gunter, Dan and Skinner, David and Ceder, Gerbrand and others},
  journal={APL materials},
  volume={1},
  number={1},
  year={2013},
  publisher={AIP Publishing}
}

@article{horton2025accelerated,
  title={Accelerated data-driven materials science with the Materials Project},
  author={Horton, Matthew K and Huck, Patrick and Yang, Ruo Xi and Munro, Jason M and Dwaraknath, Shyam and Ganose, Alex M and Kingsbury, Ryan S and Wen, Mingjian and Shen, Jimmy X and Mathis, Tyler S and others},
  journal={Nature Materials},
  volume={24},
  number={10},
  pages={1522--1532},
  year={2025},
  publisher={Nature Publishing Group UK London}
}

@article{veril2021questdb,
  title={QUESTDB: A database of highly accurate excitation energies for the electronic structure community},
  author={V{\'e}ril, Micka{\"e}l and Scemama, Anthony and Caffarel, Michel and Lipparini, Filippo and Boggio-Pasqua, Martial and Jacquemin, Denis and Loos, Pierre-Fran{\c{c}}ois},
  journal={Wiley Interdisciplinary Reviews: Computational Molecular Science},
  volume={11},
  number={5},
  pages={e1517},
  year={2021},
  publisher={Wiley Online Library}
}

@article{fan2022transition,
  title={Transition metal dichalcogenides (tmdcs) heterostructures: synthesis, excitons and photoelectric properties},
  author={Fan, Jianuo and Sun, Mengtao},
  journal={The Chemical Record},
  volume={22},
  number={6},
  pages={e202100313},
  year={2022},
  publisher={Wiley Online Library}
}

@article{farkous2020strain,
  title={Strain effects on the electronic and optical properties of Van der Waals heterostructure MoS2/WS2: a first-principles study},
  author={Farkous, M and Bikerouin, M and Thuan, Doan V and Benhouria, Y and El-Yadri, M and Feddi, E and Erguig, H and Dujardin, F and Nguyen, Chuong V and Hieu, Nguyen V and others},
  journal={Physica E: Low-dimensional Systems and Nanostructures},
  volume={116},
  pages={113799},
  year={2020},
  publisher={Elsevier}
}

@article{yang2025exploring,
  title={Exploring the frontier of 2D materials: Strain and electric field effects in MoS2/WS2 vdW heterostructures},
  author={Yang, Shang-Hsiu and Murugan, Sandhiya and Sivakumar, Chandrasekar and Hsu, Yu-Chun and Balraj, Babu and Tsia, Jia-Hau and Chen, Mei-Hsin and Ho, Mon-Shu},
  journal={Journal of Alloys and Compounds},
  volume={1012},
  pages={178457},
  year={2025},
  publisher={Elsevier}
}

@article{ghosh2020machine,
  title={Machine learning study of magnetism in uranium-based compounds},
  author={Ghosh, Ayana and Ronning, Filip and Nakhmanson, Serge M and Zhu, Jian-Xin},
  journal={Physical Review Materials},
  volume={4},
  number={6},
  pages={064414},
  year={2020},
  publisher={APS}
}


\end{document}